\documentclass[12pt]{iopart}
\usepackage{epsfig,latexsym,graphicx}
\begin{document}

\title[EXAFS study of Sm$_{0.2}$Pr$_{0.3}$Sr$_{0.5}$MnO$_{3}$]{Local structural changes in paramagnetic and charge ordered phases of Sm$_{0.2}$Pr$_{0.3}$Sr$_{0.5}$MnO$_{3}$: An EXAFS Study}
\author{K R Priolkar\dag\footnote[3]{Author to whom correspondence should be addressed.}, Vishwajeet Kulkarni\dag P R Sarode\dag and S Emura\ddag}
\address{\dag\ Department of Physics, Goa University, Goa 403 206 India.}
\address{\ddag\ Institute of Scientific and Industrial Research, Osaka University,Mihoga-oka 8-1, Ibaraki, Osaka 567-0047, Japan.}

\ead{krp@unigoa.ac.in}
\date{\today}

\begin{abstract}
Sm$_{0.5-x}$Pr$_{x}$Sr$_{0.5}$MnO$_{3}$ exhibits variety of ground states as $x$ is varied from 0 to 0.5. At an intermediate doping of $x = 0.3$ a charge-ordered CE type antiferromagnetic insulating (AFI) ground state is seen. The transition to this ground state is from a paramagnetic insulating (PMI) phase through a ferromagnetic metallic phase (FMM). Local structures in PMI and AFI phases of x = 0.3 sample  have been investigated using Pr K-edge and Sm K-edge Extended X-ray Absorption Fine Structure (EXAFS). It can be seen that the tilting and rotation of the MnO$_6$ octahedra about the b-axis are responsible for the charge ordered CE-type antiferromagnetic ground state at low temperatures. In addition a shift in the position of the rare earth ion along the c-axis has to be considered to account for observed distribution of bond distances around the rare earth ion.
\end{abstract}
\pacs{75.47.Lx, 87.64.Gb, 75.47.Gk}
\submitto{\JPCM}
\maketitle

\section{Introduction}
Colossal Magneto-Resistive (CMR) materials have provided an opportunity to understand the intricate relationship between their structural, electronic and magnetic degrees of freedom. The magnetic order in these materials is essentially ferromagnetic which is due to parallel alignment of the spin of the hopping $e_g$ electron between Mn$^{3+}$-O-Mn$^{4+}$ pairs due to double exchange mechanism \cite{zen}. Study of these materials becomes more complex at low temperatures owing to the rich magnetic, electronic and structural phases exhibited by them. Recent finding of coexistence of ferroelectricity and magnetism in manganites has only added to the existing complexity \cite{cheong,cnr,kho}. Khomskii and coworkers \cite{kho2} have pointed out that coupling between magnetic and charge-ordering in charge-ordered and orbital-ordered perovskites can give rise to ferroelectric magnetism. Although, the CMR properties have been investigated extensively, the physics behind their varied low temperature ground states is still unclear. The search for a microscopic picture of the ground state especially in half-doped manganites remains a very active research field.

Compounds belonging to the series Sm$_{0.5-x}$Pr$_{x}$Sr$_{0.5}$MnO$_{3}$ ($0 \le x \le 0.5$) exhibit significantly different low temperature ground states with $x$ and provide an opportunity to understand the causes for attaining such varied ground states. In this series, end member Sm$_{0.5}$Sr$_{0.5}$MnO$_3$ undergoes paramagnetic-insulating (PMI) to ferromagnetic-metallic (FMM) phase transition below 95 K \cite{FDam,KMGu,CMar}. The other end member Pr$_{0.5}$Sr$_{0.5}$MnO$_{3}$ exhibits PMI phase at RT, FMM phase below 255 K and a orbitally ordered antiferromagnetic-insulating (AFI) phase below 140 K \cite{KKni,YTom}. However, at intermediate doping level of $x = 0.3$, Pr$_{0.3}$Sm$_{0.2}$Sr$_{0.5}$MnO$_{3}$ exhibits CE type charge ordered (CO) state below 155 K starting from PMI phase at RT and  FMM phase in the temperature range 215 K to 155 K \cite{FDam}. Presence of CO state, for the intermediate doping, evokes interest in understanding the role of rare-earth ions in inducing the CO ground state in the material.

Role of the weighted average radius ($<r_{A}>$) of the A-site cation and the size-mismatch of A-site cations is well known in CO materials \cite{FDam,FDa2,Anth,CNRR,YTok,VNSm}. Smaller A-site cation size is reported to support charge-ordered state due to the smaller e$_g$ electron bandwidth (W) whereas greater size mismatch in A-site cations has been reported to destroy charge-ordering  and render the manganite ferromagnetic \cite{FDa2}. In all these investigations, the role of  Mn-O-Mn bond angle and also Mn-O bond distance have been the centre of focus in deciding the ground state of these materials. Ferromagnetic double-exchange, antiferromagnetic superexchange, Jahn-Teller (JT) distortion of manganese octahedra and charge-ordered state invariably depend upon these quantities.  The multiferroic properties of manganites have also been attributed to tilting of polyhedra or frustrated magnetism \cite{feibig,kimura,hur}. Some studies on charge-ordered samples have also indicated the role of covalent mixing of O(2p$_\sigma$) and A-site cations, on T$_C$ and T$_N$ \cite{CNRR,CNR2}. It has been argued that the covalent mixing of the anion O(2p$_\sigma$) orbital with the A-site cation 4f orbitals, competing against its covalent mixing with the Mn 3d$e_g$ orbital, plays an important role in deciding the properties of manganites. Mizokawa et al have also expressed the possibility of a 4f-O(2p$_\pi$) band to accommodate some holes leading to reduction in the effective hole concentration in the Mn 3d-O(2p$\sigma$)  band \cite{TMiz}.

Hence, local structure around Mn sites needs to be probed carefully to understand the role of MnO$_6$ octahedra in various phases exhibited by such samples. Such investigations using Mn K-edge EXAFS have been undertaken earlier in some CMR materials \cite{CMen,CHBo,GSub,Tomo,DCao}. However, similar Mn K-edge EXAFS study on Sm$_{0.2}$Pr$_{0.3}$Sr$_{0.5}$MnO$_{3}$ is not possible  as Mn K-edge EXAFS is severely affected by EXAFS of Pr/Sm  L-edges. Recently it has been shown that \cite{KRPr} that information about MnO$_6$ octahedra could be deduced from the study of rare-earth K-edge EXAFS. Hence it will be interesting to investigate the local structure of A-site cations  in various phases of a CO compound in order to glean an insight on the role of MnO$_6$ octahedra in stabilizing a particular ground state.  We report herein EXAFS of A-site cation (Pr/Sm) in PMI and CE-type AFI charge-ordered phases of Sm$_{0.2}$Pr$_{0.3}$Sr$_{0.5}$MnO$_{3}$.

\section{Experimental}
Polycrystalline Sm$_{0.2}$Pr$_{0.3}$Sr$_{0.5}$MnO$_{3}$ sample was synthesized using  the conventional solid-state reaction method using the preparation technique mentioned in \cite{FDam}. The samples were characterized  by XRD, resistivity and AC susceptibility measurements. X-ray  powder  diffraction patterns were recorded  with Cu K$\alpha$ radiation ($\lambda$ = 1.5418\AA) at  room temperature between $2\theta$  = 20$^\circ$ to 100$^\circ$ with a step 0.02$^\circ$ on a Siemens D5000 diffractometer. Low field a.c. susceptibility measurements were carried out during warming from liquid nitrogen temperature to 300 K. Standard four-probe method was used to measure resistivity of the samples in the temperature interval of 80K to 300K. EXAFS spectra at Pr and Sm K-edges were recorded using BL01B1 XAFS beam-line at SPring-8. Si(311) crystal plane served as the monochromator. The first mirror and the monochromator were fully tuned in order to get optimal resolution. The slit width of the monochromator exit was 0.3 mm vertical and 6 mm horizontal to ensure good signal to noise ratio. The photon energy  was calibrated for each scan with the first inflection point of Pr K-edge in Pr metal (42002 eV). Both the incident (I$_0$) and the transmitted (I) synchrotron beam intensities were measured simultaneously using ionization chambers filled with a mixture of 15\% Ar and 85\% N$_2$ gases and 100\% Ar gas respectively. The absorbers were made by pressing the samples into pellets of 10 mm diameter with boron nitride as binder. The thickness of the absorber was adjusted such that $\Delta\mu x$ was restricted to a value in the range 0.5 to 1, where $\Delta\mu$ is edge step in the absorption coefficient and $x$ is the sample thickness. Data analysis was done using the ATHENA and ARTEMIS \cite{MNew}.

\section{Results and Discussion}
\begin{figure}
 \centering
 \epsfig{file=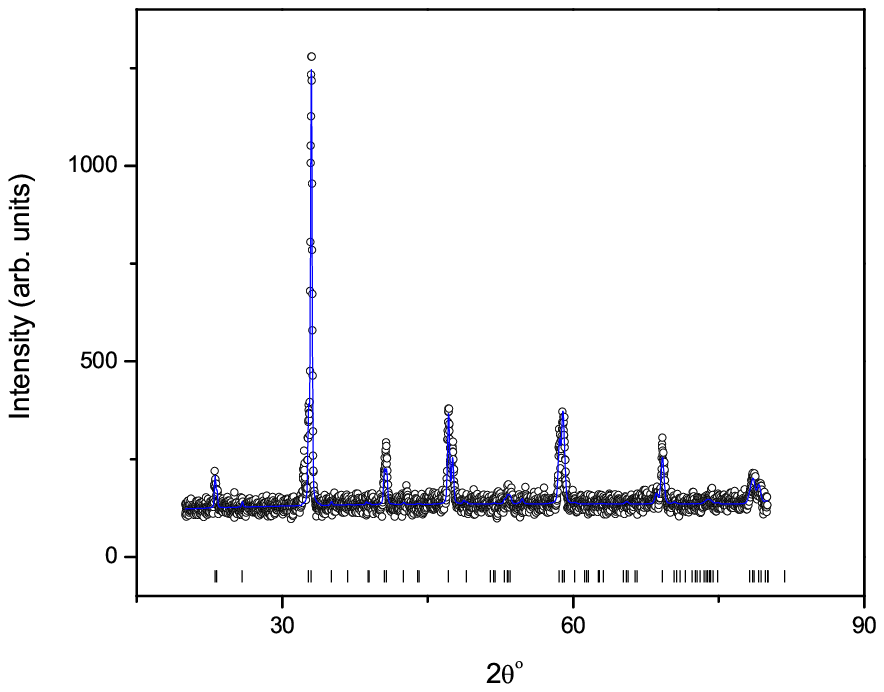, width=8cm, height=6cm}
\caption{\label{fig:XRD} Room temperature XRD pattern of Sm$_{0.2}$Pr$_{0.3}$Sr$_{0.5}$MnO$_{3}$. The solid line
is the calculated pattern for Pbnm structure and the markers below indicate positions of Bragg reflections.}
\end{figure}

\begin{figure}
  \centering
 \epsfig{file=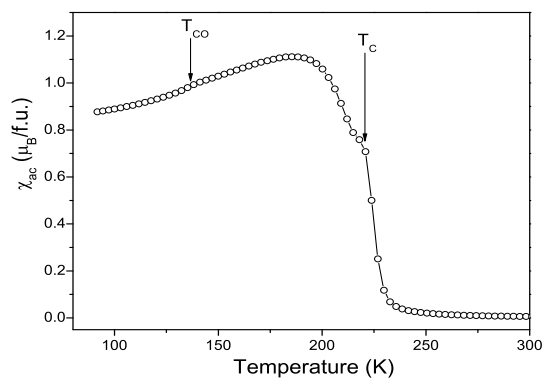, width=8cm, height=6cm}
\caption {\label{fig:SUS} AC susceptibility versus temperature plot of Sm$_{0.2}$Pr$_{0.3}$Sr$_{0.5}$MnO$_{3}$. Vertical arrows indicate paramagnetic to ferromagnetic and charge ordering transitions.}
\end{figure}

Room temperature XRD patterns of the sample, presented in Fig.\ref{fig:XRD}, shows primarily a single phase sample. No crystallographic data is available in literature on this sample. However, the diffraction pattern is very similar to that of Sm$_{0.5}$Sr$_{0.5}$MnO$_3$ which is reported to have orthorhombic structure \cite{kurb}. Indexing our diffraction pattern with Pnma space group identifies all the reflections and gives lattice parameters as $a$ = 5.4301 \AA, $b$ = 7.6471 \AA, $c$ = 5.4733 \AA. A.c. susceptibility as a function of temperature presented in Fig.\ref{fig:SUS} and the four-probe resistivity measurements (not shown) performed on these samples closely match with those reported in literature \cite{FDam}.

\begin{figure}
 \centering
 \epsfig{file=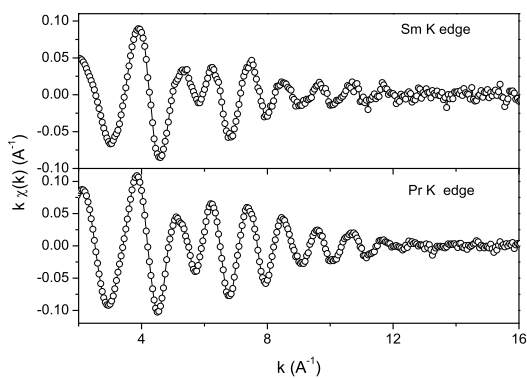, width=8cm, height=6cm}
\caption{\label{fig:chik} k weighted room temperature (300K) EXAFS data plots at Pr and Sm K-edge in Sm$_{0.2}$Pr$_{0.3}$Sr$_{0.5}$MnO$_{3}$.}
\end{figure}

\begin{figure}
 \centering
 \epsfig{file=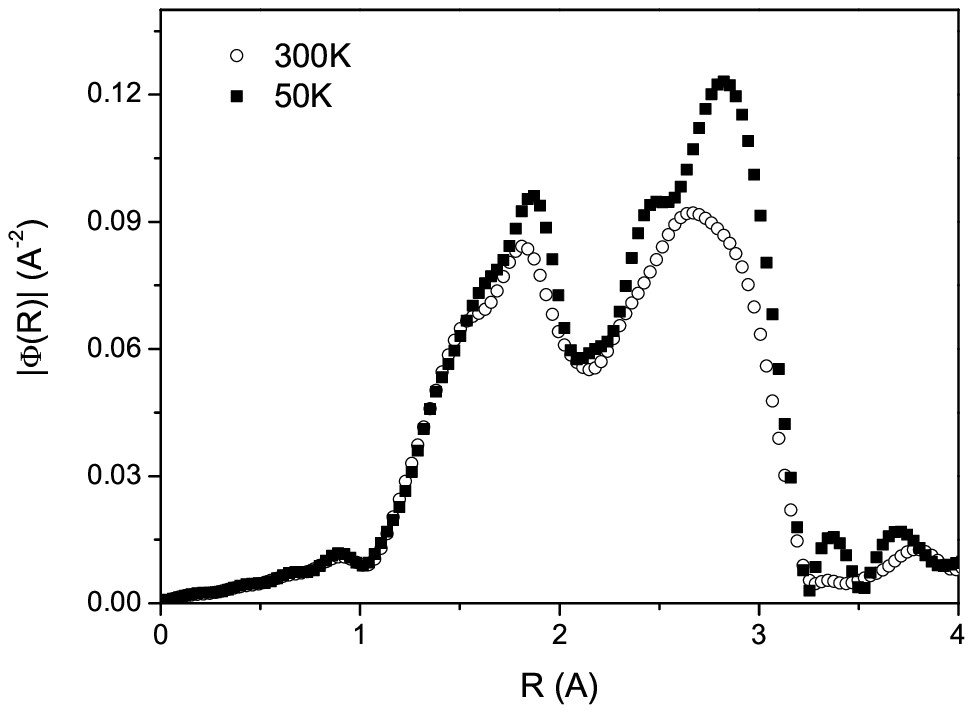, width=8cm, height=6cm}
\caption{\label{fig:fouripr} Magnitude of Fourier transforms of Pr K-edge EXAFS in  Sm$_{0.2}$Pr$_{0.3}$Sr$_{0.5}$MnO$_{3}$ at 300K and 50 K.}
\end{figure}

Weighted EXAFS spectra, $k \chi(k)$ versus $k$ plots of Sm$_{0.2}$Pr$_{0.3}$Sr$_{0.5}$MnO$_{3}$  at room temperature (R.T.) at Pr and Sm K edges are presented in Fig.\ref{fig:chik}. It can be seen that EXAFS oscillations are visible upto 16 \AA$^{-1}$ indicating usable data range. Fourier Transforms (FT) (not corrected for phase shift) of Pr K-edge EXAFS in Pr$_{0.3}$Sm$_{0.2}$Sr$_{0.5}$MnO$_{3}$ at R. T. and 50K are presented in Fig.\ref{fig:fouripr}. New features appear at around 2.4 \AA~ and 3.25 \AA~ at  50 K as compared to that at room temperature. This is perhaps due to redistribution of Pr-O coordination numbers at 50 K. In addition, a slight shift of the main peak in FT towards higher R-side indicates longer Pr-Mn bonds at 50 K. It reflects a gradual increase in  Pr-Mn bond lengths from RT to 50 K. 

\begin{figure}
\centering
\epsfig{file=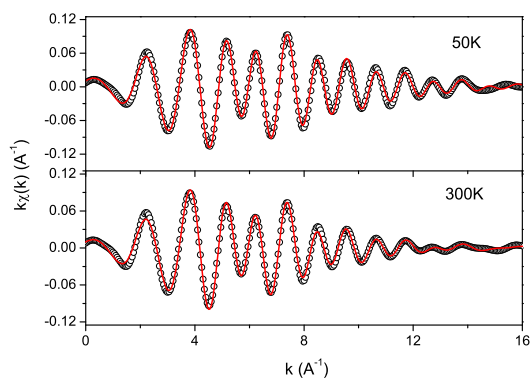, width=8cm, height=6cm} 
\caption{\label{fig:prqfits} Back transformed Pr K-edge  EXAFS in  Sm$_{0.2}$Pr$_{0.3}$Sr$_{0.5}$MnO$_{3}$ in k-space. Circles represent experimental data and line represents the best fit.}
\end{figure}

\begin{figure}
 \centering
 \epsfig{file=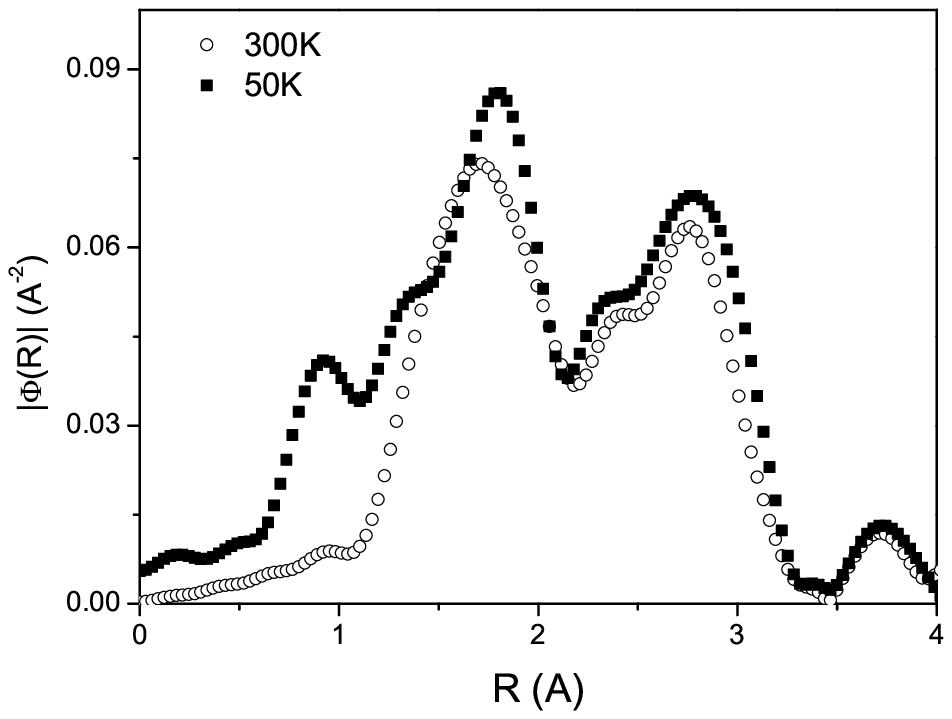, width=8cm, height=6cm}
\caption{\label{fig:foursm1} Magnitude of Fourier transforms of Sm K-edge EXAFS in  Sm$_{0.2}$Pr$_{0.3}$Sr$_{0.5}$MnO$_{3}$ at 300K and 50 K.}
\end{figure}

\begin{figure}
\centering
\epsfig{file=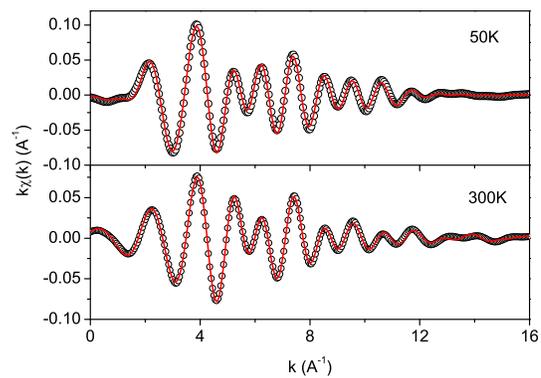, width=8cm, height=6cm} 
\caption{ \label{fig:sm1qfits} Back transformed Sm K-edge EXAFS in  Sm$_{0.2}$Pr$_{0.3}$Sr$_{0.5}$MnO$_{3}$ in k-space. Circles represent experimental data and the best fit to the data is shown as line.}
\end{figure}

Pr/Sm K-edge EXAFS were fitted with k$^1$ weighting in k-range of 2 to 16 \AA$^{-1}$ and in the R-range between 1 to 3.5 \AA~ to include Pr/Sm-O and Pr/Sm-Mn bond lengths. As no structural report is available for this compound, the theoretical EXAFS function ($\chi(k)$) including amplitude and phase, coordination numbers and  bond distances (given as calculated bond lengths in Tables \ref{prexafs} and \ref{smexafs}) were calculated using the crystallographic data available for isostructural and charge ordered compound, Nd$_{0.5}$Sr$_{0.5}$MnO$_3$ \cite{PGRa} with lattice parameters obtained from XRD. The value of  k$_{max}$ = 16 \AA$^{-1}$ gives a resolution in R-space of about 0.1\AA. Therefore all the Pr/Sm-O bond lengths with difference less than 0.1\AA~ were grouped together. For example, Pr has one oxygen neighbour at 2.418\AA~ (written as 2.418 $\times$ 1) and 2 oxygens at 2.443\AA. These were clubbed togather as one bond distance at 2.43\AA~ with a coordination number 3. Thus three Pr-O bond lengths at 2.43\AA, 2.69\AA~ and 3.06\AA~ with coordination numbers of 3, 6 and 3 respectively and two Pr-Mn bond lengths at 3.23\AA~ and 3.35\AA~ with coordination number 2 and 6 respectively were realized. Using these constraints the theoretical EXAFS function was fitted to the experimental data with the bond length and mean square relative dispalcement ($\sigma^2$) along with amplitude reduction factor and $\Delta E_0$ as variable parameters. The coordination numbers was kept fixed to the above distribution. The final fitted parameters for  Pr K-edge EXAFS are presented in Table \ref{prexafs} and fitting in k-space is shown in Fig. \ref{fig:prqfits}. It can be seen that the one of the three Pr-O bond length obtained from EXAFS analysis is much longer than the one calculated from XRD. It must be noted here that one obtains an average picture from XRD and hence local distortions in the structure are masked in comparison with EXAFS which is essentially a local probe. Here, since there are three cations (Pr, Sm and Sr) occupying the A site, such a distortion could be expected. The EXAFS at 50 K cannot be fitted with the above distribution of bond lengths and coordination numbers. Different combinations of coordination numbers were tried and the best fits were obtained for two cases: one with coordination number of 4 for all three Pr-O and two Pr-Mn bond lengths and second one with 6, 3 and 3 combination for three Pr-O bond lengths and 4 + 4 combination for Pr-Mn bond lengths. The second model was rejected as it gave negative $\sigma^2$ for one of the Sm-O bond length when it was applied to Sm K EXAFS. Fourier Transforms of Sm K-edge EXAFS of Sm$_{0.2}$Pr$_{0.3}$Sr$_{0.5}$MnO$_{3}$  and their fitting in k-space are presented in Fig. \ref{fig:foursm1} and Fig. \ref{fig:sm1qfits} respectively.  Sm-O and Sm-Mn bond lengths are relatively smaller compared to the corresponding  Pr-O and Pr-Mn bond lengths [Table \ref{smexafs}], due to the smaller ionic radius of Sm ion. However, the changes in Sm-O and Sm-Mn distances in going from 300K to 50K are similar to those in Pr-O and Pr-Mn respectively. Furthermore, it may also be noted from Table \ref{prexafs} and \ref{smexafs} that there is an overall contraction of Pr/Sm-O bonds and expansion of Pr/Sm-Mn bonds as the system evolves from PMI (300K) to CO (50K) state.

\begin{table}
\caption {\label{prexafs} Structural parameters of Pr ion in Sm$_{0.2}$Pr$_{0.3}$Sr$_{0.5}$MnO$_{3}$.  Figures in
brackets indicate  uncertainty  in the last digit.}
\begin{indented}
\item
\begin{tabular}{@{}lccccc}
  \br
& &\multicolumn{2}{c}{300K} & \multicolumn{2}{c}{50K}\\
 \mr
  Bond & \multicolumn{2}{c}{Bond Length $\times$ CN$^a$} & $\sigma^2$ & Bond Length $\times$ CN$^a$ & $\sigma^2$\\
&  calculated (\AA) & fitted  (\AA) & (\AA)$^{2}$ & fitted  (\AA) & (\AA)$^{2}$\\
\mr
Pr-O & 2.418 $\times$ 1 & 2.432(1) $\times$ 3 & 0.007(2) & 2.423(8)$\times$ 4 & 0.010(1)\\
& 2.443 $\times$ 2 & & & &\\
\mr
Pr-O & 2.591 $\times$ 1 & & & &\\
& 2.633 $\times$ 2 & 2.639(2) $\times$ 6 & 0.023(4)& 2.602(1)$\times$ 4 & 0.013(2)\\
& 2.716 $\times$ 2 & & & &\\
& 2.881 $\times$ 1 & & & &\\
\mr
Pr-O & 3.026 $\times$ 1 & 3.270(2) $\times$ 3 & 0.002(1) & 3.320(2) $\times$ 4 & 0.004(1)\\
& 3.085 $\times$ 2 & &  & &\\
\mr
Pr-Mn & 3.233 $\times$ 2 & 3.266(1) $\times$ 2& 0.004(1) &  3.285(1) $\times$ 4 & 0.003(5)\\
\mr
Pr-Mn & 3.303 $\times$ 2 & & & & \\
& 3.346 $\times$ 2 & 3.373(5) $\times$ 6 & 0.006(6)& 3.374(5) $\times$ 4 & 0.002(6)\\
& 3.406 $\times$ 2 & & & & \\
\br
 $^a$ Coordination Number 
\end{tabular}
\end{indented}
\end{table}

\begin{table}
\caption {\label{smexafs} Structural parameters of Sm ion in Sm$_{0.2}$Pr$_{0.3}$Sr$_{0.5}$MnO$_{3}$. Figures in
brackets indicate  uncertainty the last digit.}
\begin{indented}
 \item \begin{tabular}{@{}lccccc}
\br
& &\multicolumn{2}{c}{300K} & \multicolumn{2}{c}{50K}\\
\mr
Bond & \multicolumn{2}{c}{Bond Length $\times$ CN$^a$} & $\sigma^2$ & Bond Length $\times$ CN$^a$ & $\sigma^2$\\
&  calculated (\AA) & fitted  (\AA) & (\AA)$^{2}$ & fitted  (\AA) & (\AA)$^{2}$\\
\mr
Sm-O & 2.418 $\times$ 1 & 2.415(1) $\times$ 3 & 0.009(2) & 2.446(8)$\times$ 4 & 0.008(1)\\
& 2.443 $\times$ 2 & & \\
\mr
Sm-O & 2.591 $\times$ 1 & & \\
& 2.633 $\times$ 2 & 2.645(2) $\times$ 6 & 0.033(4) & 2.579(1)$\times$ 4 & 0.033(2)\\
& 2.716 $\times$ 2 & &\\
& 2.881 $\times$ 1 & & \\
\mr
Sm-O & 3.026 $\times$ 1 & 3.103(2) $\times$ 3 & 0.007(1) & 3.298(2) $\times$ 4 & 0.004(1)\\
& 3.085 $\times$ 2 & & \\
\mr
Sm-Mn & 3.233 $\times$ 2 & 3.272(1) $\times$ 2& 0.004(1) & 3.267(1) $\times$ 4 & 0.011(5)\\
\mr
Sm-Mn& 3.303 $\times$ 2 & & \\
& 3.346 $\times$ 2 & 3.339(5) $\times$ 6 & 0.009(6) & 3.343(5) $\times$ 4 & 0.0055(6)\\
& 3.406 $\times$ 2 & & \\
\br
 $^a$ Coordination Number
\end{tabular}
\end{indented}
\end{table}

\begin{figure}
\centering
 \epsfig{file=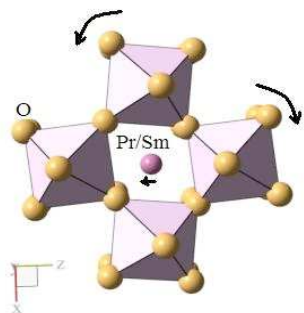, width=6cm, height=8cm}
  \caption{\label{prsmUNIT} Local Structure around Pr/Sm ion at room temperature and at 50 K in Sm$_{0.2}$Pr$_{0.3}$Sr$_{0.5}$MnO$_{3}$.}
\end{figure}

Local structure investigations on this material  are modeled on orthorhombic structure in space group Pnma. In going from PMI to CO state most of the Pr/Sm-O and Pr/Sm-Mn bond-lengths remain nearly similar. The changes happen in the coordination numbers which change from 3 + 6 + 3 to 4 + 4 + 4 for Pr/Sm-O bond lengths and from 2 + 6 to 4 + 4 for Pr/Sm-Mn bond lengths. The Pr/Sm-O bond length at about 2.6\AA~ at 300K is made up of 4 FEFF calculated Pr/Sm-O bond lengths spread from 2.59\AA~ to 2.88\AA. This wide distribution in the bond distances is reflected in the corresponding value of $\sigma^2$ which is more than 0.02\AA.  The changes in Pr/Sm-O coordination numbers at RT and 50 K [Tables \ref{prexafs} and \ref{smexafs}] can be understood in terms of rotation and tilting of Mn-O octahedra along b-axis.  In charge-ordered state, the neighboring MnO$_6$ octahedra rotate about b-axis in opposite directions as shown in Fig. \ref{prsmUNIT}. This cooperative rotation effect affects mainly the Pr/Sm-O distances in the $ab$ and $bc$ planes. In effect one of the six bond lengths at 2.6\AA~ shortens to 2.44\AA~ while another one in the diagonally opposite plane elongates to $\approx$ 3.25\AA. At room temperature, oxygen ions in Mn-O-Mn chain along b-axis are off-axis resulting  in Mn-O-Mn bond angle of $\sim$ 174$^\circ$. This Mn-O-Mn angle decreases in CO state due to tilting of the octahedra and results in shortening of one of the Pr/Sm-O bonds and lengthening of the other Pr/Sm-O bond. The reduced Mn-O-Mn angle also decreases the double exchange transport in the Mn-O-Mn chain rendering the sample insulating.

In the CO phase the crystal structure is reported to belong to P2$_1$/n space group\cite{PGRa}. The Pr/Sm-O and Pr/Sm-Mn bond lengths generated from this space group can be grouped into 6 + 3 + 3 and 2 + 6 respectively. It has been already mentioned that 6 + 3 + 3 distribution of Sm-O bond lengths resulted in negative $\sigma^2$ or a much higher R-factor of fitting. However, even with this distribution of oxygens around the rare-earth ion, Pr/Sm-Mn coordination numbers had to be 4 + 4 and not 2 + 6. This change in the distribution of Mn ions around the rare-earth can only be reconciled if one assumes a shift of the rare-earth or the Mn ion along c-axis. Shifting of the rare-earth ion  along c-axis decreases two Pr/Sm-Mn bond lengths from 3.36\AA~ to 3.28\AA. Such a shift of cations has been reported in CE-type charge-ordered compounds \cite{PGRa, JBla}.

The cooperative rotation of Mn-O octahedra observed here indirectly from the changes in Pr/Sm-O bond lengths also hint towards JT distortions. JT distortion results in unequal distribution of Mn-O bond lenghts in Mn$^{3+}$O$_6$ octahedra.  The rotation of MnO$_6$ octahedra along the b-axis, moves one of the oxygen ion closer to a particular Mn site, simultaneously increasing its distance from another Mn ion in the neighboring octahedra. Such type of unequal bond distribution in neighboring MnO$_{6}$ octahedra results in JT distortions and is known in CE type charge-ordered materials \cite{PGRa, JBla}.

The overlap between O(2p$_{\sigma}$) and 4f orbitals of rare-earth cations is one of probable causes for inducing charge-ordered state in manganites \cite{CNRR, TMiz}. In the present case, shifting of RE cations increases the overlap between 4f orbitals of rare-earth cation and O(2p$_{\sigma}$) orbitals. Such an overlap reduces itinerant electron density in the e$_g$ band. Furthermore, tilting and rotation of the MnO$_6$ octahedra reduce the Mn-O-Mn angle, thereby decreasing  the ferromagentic double exchange transfer in Mn-O-Mn chains. As a combined result, with 3d$_{(x^2-y^2)}$ orbitals depleted of electron density and rotation of MnO$_6$ octahedra the ferromagnetic double exchange between Mn$^{3+}$-O-Mn$^{4+}$ Zener pairs is drastically reduced. Moreover, unequal Mn-O distance in O-Mn-O chain strengthens the antiferromagnetic superexchange between the t$_{2g}$ electrons. Hence, shifting of RE ion and rotation together with tilting of MnO$_6$ octahedra drive the material into CE type AFI state at low temperatures.

\section{Conclusion}
We have investigated the local structure of MnO$_6$ octahedra in Sm$_{0.2}$Pr$_{0.3}$Sr$_{0.5}$MnO$_{3}$ sample with  the help of Pr and Sm K-edge EXAFS. In this case the transition to CO ground state is strongly associated with local structural environment of A-site cation. CO state in Sm$_{0.2}$Pr$_{0.3}$Sr$_{0.5}$MnO$_{3}$ arises from the combined effect of  rotation and tilting of MnO$_6$ octahedra and shifting of the RE cations. Rotation and tilting of MnO$_6$ octahedra leads to reduced Mn-O-Mn angle and shifting of RE cations increases the overlap between O$_{2p\sigma}$ and 4f orbitals of A-site cations thereby decreasing the carrier density in the e$_g$ level. This effect reduces ferromagnetic exchange between Zener pairs and strengthens antiferromagnetic superexchange interactions. Trapping of e$_g$ electrons which can  be viewed as the JT polaronic effect that leads to the charge ordering state.

\ackn Authors [KRP and PRS] would like to gratefully acknowledge JASRI for local hospitality and beam time on BL01B1 and travel assistance from Department of Science and Technology, Government of India, New Delhi. KRP thanks DST for financial support under DST Fast Track Scheme for Young Scientists (PS-19).

\newpage

\end{document}